\documentclass[aps,prb,twocolumn,superscriptaddress,showpacs,preprintnumbers,amsmath,amssymb]{revtex4}

\usepackage{graphicx}
\usepackage{dcolumn}
\usepackage{bm}
\usepackage{hyperref}
\usepackage[latin1]{inputenc}
\usepackage{xspace}
\usepackage{xfrac}				
				
\newcommand{\OX}{La$_2$O$_2$Fe$_2$OSe$_2$\xspace}
\newcommand{\NdOFeSe}{Nd$_2$O$_2$Fe$_2$OSe$_2$\xspace}
\newcommand{\LaOFeSe}{La$_2$O$_2$Fe$_2$OSe$_2$\xspace}

\newcommand{\SrFFeS}{Sr$_2$F$_2$Fe$_2$OS$_2$\xspace}
\newcommand{\BaFFeSe}{Ba$_2$F$_2$Fe$_2$OSe$_2$\xspace}
\newcommand{\LaOMnSe}{La$_2$O$_2$Mn$_2$OSe$_2$\xspace}
\newcommand{\LaOCoSe}{La$_2$O$_2$Co$_2$OSe$_2$\xspace}

\begin{document}

\title{Magnetic order and spin dynamics in \OX probed by $^{57}$Fe M\"ossbauer,\\ $^{139}$La NMR, and muon spin relaxation spectroscopy} 

\author{M. G\"unther}
\email{guenth42@googlemail.com}
\author{S. Kamusella}
\author{R. Sarkar}
\author{T. Goltz}
\affiliation{Institut f\"ur Festk\"orperphysik, Technische Universit\"at Dresden, D-01069 Dresden, Germany}
\author{H. Luetkens}
\author{G. Pascua}
\affiliation{Paul Scherrer Institut, 5232 Villigen PSI, Switzerland}
\author{S.-H. Do}
\author{K.-Y. Choi}
\affiliation{Department of Physics, Chung-Ang University, Seoul 156-756,  Republic of Korea  }
\author{H.~D.~Zhou}
\affiliation{Department of Physics and Astronomy, University of Tennessee, Knoxville, Tennessee 37996-1200, USA}
\affiliation{National High Magnetic Field Laboratory, Florida State University, Tallahassee, FL 32306-4005, USA}
\author{C.G.F. Blum}
\affiliation{Leibniz-Institut für Festk\"orper- und Werkstoffforschung (IFW) Dresden, D-01171 Dresden, Germany}
\author{S. Wurmehl}
\affiliation{Leibniz-Institut für Festk\"orper- und Werkstoffforschung (IFW) Dresden, D-01171 Dresden, Germany}
\affiliation{Institut f\"ur Festk\"orperphysik, Technische Universit\"at Dresden, D-01069 Dresden, Germany}
\author{B. B\"uchner}
\affiliation{Leibniz-Institut für Festk\"orper- und Werkstoffforschung (IFW) Dresden, D-01171 Dresden, Germany}
\affiliation{Institut f\"ur Festk\"orperphysik, Technische Universit\"at Dresden, D-01069 Dresden, Germany}
\author{H.-H. Klauss}
\affiliation{Institut f\"ur Festk\"orperphysik, Technische Universit\"at Dresden, D-01069 Dresden, Germany}

\date{\today}

\begin{abstract}
We present a detailed local probe study of the magnetic order in the oxychalcogenide
 \OX utilizing $^{57}$Fe M\"ossbauer, $^{139}$La NMR, and muon spin relaxation spectroscopy.
 This system can be regarded as an insulating reference system of the Fe arsenide and
 chalcogenide superconductors.
From the combination of the local probe techniques we identify a non-collinear
magnetic structure similar to \SrFFeS. 
The analysis of the magnetic order parameter yields an ordering temperature $T_\mathrm{N}~=~90.1~\mathrm{K}$ and 
a critical exponent of $\beta=0.133$, which is close to the 2D Ising universality class as reported in the related oxychalcogenide family.

\end{abstract}


\maketitle
\section{Introduction}
One of the important issues to understand the iron based superconductors is the role of
electron correlations. 
The antiferromagnetically (AFM)
ordered state of the parent compounds of \emph{R}FeAsO (\emph{R} = rare earth)\cite{deLaCruz-Nature-2008,Klauss-PRL-2008} and BaFe$_2$As$_2$ \cite{Rotter-PRB-2008} 
pnictide superconductors can be understood either as a
spin density wave order of an itinerant multiband Fermi surface,\cite{Mazin-Nature-2010} or due to a
frustrated $J_1$-$J_2$ interaction of localized spins.\cite{Si-NatPhys-2009,Fernandes-PRB-2012}
Strongly localized systems with
layered transition metal (\emph{Tm}) pnictide or (oxy-) chalcogenide structures serve as reference systems to
understand the properties of the iron based superconductors.
Moreover, due to the interplay of three different intraplane exchange interactions of \emph{Tm} ions,
the oxychalcogenide systems show a rich variety of magnetically ordered ground states.
G-type AFM order with magnetic Mn moments aligned along the crystallographic $c$-axis is reported for \LaOMnSe.\cite{Ni-2010-PRB}
A non-collinear arrangement of magnetic moments parallel to the $a$-$b$ plane is concluded for \SrFFeS 
[Fig.~\ref{fig:Structure}(e)] and a different one for \LaOCoSe 
[with a possible magnetic structure according to the model in Fig.~\ref{fig:Structure}(d)]. \cite{Zhao-PRB-2013,Fuwa-JACS-2010} 

The \emph{R}$_2$O$_2$\emph{Tm}$_2$O\emph{Ch}$_2$ and \emph{A}$_2$F$_2$\emph{Tm}$_2$O\emph{Ch}$_2$ oxychalcogenides 
exist in a rich variety of compositions: systems with \emph{R}~=~La, Ce, Pr, Nd, Sm; \emph{A}~=~Sr, Ba; \emph{Tm}~=~Fe, Co, Mn; and
 \emph{Ch}~=~Se, S are reported.\cite{Mayer1992,Kabbour-JACS-2008,Fuwa-JPCM-2010,Ni-2010-PRB,Free2011,Landsgesell2013,Lei-PRB-2012} 
Among these isostructural
materials with $I4/mmm$ symmetry, all reported systems with \emph{Tm}~=~Fe exhibit a transition to antiferromagnetic 
order near $T_\mathrm{N}$~=~100~K.
Non-metallic and antiferromagnetic properties in La$_2$O$_2$Fe$_2$O\emph{Ch}$_2$ and \SrFFeS are consistent with theoretical studies
indicating Mott-insulating behavior. \cite{Zhu-2010,Wang-2011-SSC,Zhao-PRB-2013}  
 $^{57}$Fe~M\"ossbauer studies on \NdOFeSe   
and \SrFFeS found critical exponents close to the 2D Ising universality class, 
indicating anisotropical, dominantly 2D magnetic exchange interactions. \cite{Kabbour-JACS-2008,Fuwa-JPCM-2010,Zhao-PRB-2013}     

In these systems, within a Fe$_2$O\emph{Ch}$_2$ layer, three different magnetic exchange interactions of $\mathrm{Fe}^{2+}$ ions are considered:
A diagonal nearest neighbor (nn) interaction $J_{\mathrm{nn}}$ and two 
next nearest neighbor (nnn) interactions $J_{\mathrm{nnn1}}$
and $J_{\mathrm{nnn2}}$ via $\mathrm{O}^{2-}$ or two $\emph{Ch}^{2-}$ ions, respectively [Fig.~\ref{fig:Structure}(b)].\cite{Kabbour-JACS-2008} 
The nature of the magnetic interactions and possible models of magnetic order in the iron based oxychalcogenides 
had been studied extensively.\cite{Kabbour-JACS-2008,Fuwa-JPCM-2010,Free-2010-PRB,Wang-2011-SSC,Zhao-PRB-2013,Zhu-2010}

The magnetic ground state of \LaOFeSe was proposed to obey a plaquette AFM order for on-site Hubbard interaction $U=1.5$~eV or $U=3$~eV, and
the Neel state for $U=4.5$~eV. \cite{Zhu-2010} 
From this the deduced interactions are reported to be 
FM for the Fe-$\mathrm{Se}_2$-Fe pathway, AFM for the Fe-O-Fe interaction, and AFM for the nearest neighbor coupling $J_\mathrm{nn}$.
The transition from plaquette AFM order ($p>1$) to the Neel state ($p<1$) is driven by the parameter $p=J_{\mathrm{nnn1}}/|J_{\mathrm{nn}}|$.
While in the plaquette state two nn Fe magnetic moments are aligned parallel and two antiparallel, with stronger $J_\mathrm{nn}$
in the Neel state all AFM nn interactions are satisfied, but not the AFM $J_{\mathrm{nnn1}}$ mediated by oxygen ions.  

A neutron diffraction study on \LaOFeSe concluded a collinear and columnar AFM order 
below $T_\mathrm{N}=~90$~K. \cite{Free-2010-PRB}
In this model non of the three interactions is satisfied for all nn and nnn Fe moments. 
For one axis of the basal plane the Fe magnetic moments are aligned parallel along Fe-O-Fe and Fe-$\mathrm{Se}_2$-Fe chains and
antiparallel along the other.   
Again two nn Fe magnetic moments are aligned parallel and two antiparallel.   

In contrast, Fuwa et al. introduced two different non-collinear models of magnetic order for \NdOFeSe.\cite{Fuwa-JPCM-2010} 
These models are consistent with a axial anisotropy of magnetic Fe moments, aligned along perpendicular O-Fe-O chains in the 
Fe$_2$OSe$_2$ layer. 
One of these models was found to describe non-collinear magnetic order in \SrFFeS. \cite{Zhao-PRB-2013}
In Fig.~\ref{fig:Structure}(b)-(d) the three magnetic structures consistent with AFM $J_{\mathrm{nnn1}}$ interaction and
an orientation of magnetic moments along O-Fe-O chains are shown.

Recently it was shown that a collinear and columnar AFM order and the
non-collinear order concluded for \SrFFeS are indistinguishable in elastic
neutron scattering on \LaOFeSe powder, whereas inelastic experiments are only
consistent with the non-collinear model.\cite{McCabe-PRBR-2014}
Moreover, gapped ($\approx 5$~meV) and confined (below $\approx 25$~meV) magnetic 
excitation was reported.

In this study we present $^{57}$Fe M\"ossbauer, $^{139}$La NMR, and muon spin relaxation ($\mu$SR)
experiments on polycrystalline \LaOFeSe.
We find long-range magnetic order below $T_\mathrm{N}=90.1$~K and deduce a critical exponent $\beta~=~0.133$.
The magnetic order is deduced to obey a non-collinear structure according to a ferromagnetic (FM) interaction $J_{\mathrm{nnn2}}$ [Fig.~\ref{fig:Structure}(d)].
We first give a brief introduction to the experimental setup in section \ref{sec:Exp}.
In section \ref{sec:MoeSR} we present our results of $^{57}$Fe M\"ossbauer spectroscopy. 
$^{139}$La NMR field-swept and $\mu$SR spectra are presented in section \ref{sec:NMR}
and \ref{sec:muSR}.  
Finally we give a summary in section \ref{sec:Summary}.

\section{Experimental Details}
\label{sec:Exp}
Polycrystalline samples of \LaOFeSe were prepared by the solid-state reaction method 
using stoichiometric amounts of $\mathrm{La}_2\mathrm{O}_3$, Fe, and Se.  The resulting powder were ground in an argon glove box, 
pressed into a pellet, and placed inside a quartz ampoule under vacuum. The pellet was heated by following 
the routine described in the reference. \cite{Free-2010-PRB} 
The single phase of \LaOFeSe was confirmed by X-ray powder diffraction 
experiments performed on a STOE Stadi P powder diffractometer
with $\mathrm{Mo}~K_{\alpha 1}$ radiation ($\lambda=0.70926~\mathrm{Å}$) at 293~K.
The diffractometer is equipped with a curved Ge(111) monochromator and a $6^\circ$-linear
position sensitive detector.
The structural analysis was carried out by Rietveld refinement utilizing the FullProf software package.\cite{Rietveld-JApplCryst-1969,Roisnel-MatrSciForum-2001}
The refined structural parameters are presented in table \ref{table:XRDTable2} and are 
in good agreement with earlier studies. \cite{Mayer1992,Free2011}  
\begin{table} 
\begin{tabular}{c c c c c c}
\hline \hline
                                & La                & Fe             & Se           &  O(1)           &   O(2)          \\
\hline
Wykoff position                 & $4c$              & $4c$           & $4e$         &  $4d$           &    $2b$         \\
 $x (a)$                        & $\sfrac{1}{2}$    & $\sfrac{1}{2}$ & 0            &  $\sfrac{1}{2}$ & $\sfrac{1}{2}$  \\
 $y (a)$                        & $\sfrac{1}{2}$    & 0              & 0            &  0              & $\sfrac{1}{2}$  \\
 $z (c)$                        & 0.1845(1)         & 0              & 0.0966(1)    &  $\sfrac{1}{4}$ & 0               \\

\hline
\multicolumn{6}{c}{space group $I4/mmm$, $a=4.08744(4)$, $c=18.6061(3)$,} \\
\multicolumn{6}{c}{$R_w = 2.76$, $R_{wp}=3.75$, $\chi^2 =2.39$} \\
\hline
\end{tabular}
	\caption{Summary of crystallographic parameters from Rietveld refined XRD experiment.}  
	\label{table:XRDTable2}
\end{table}
All experimental work presented here were done on one single batch of crushed polycrystalline material.

\begin{figure}
	\centering
		\includegraphics[width=0.5\textwidth]{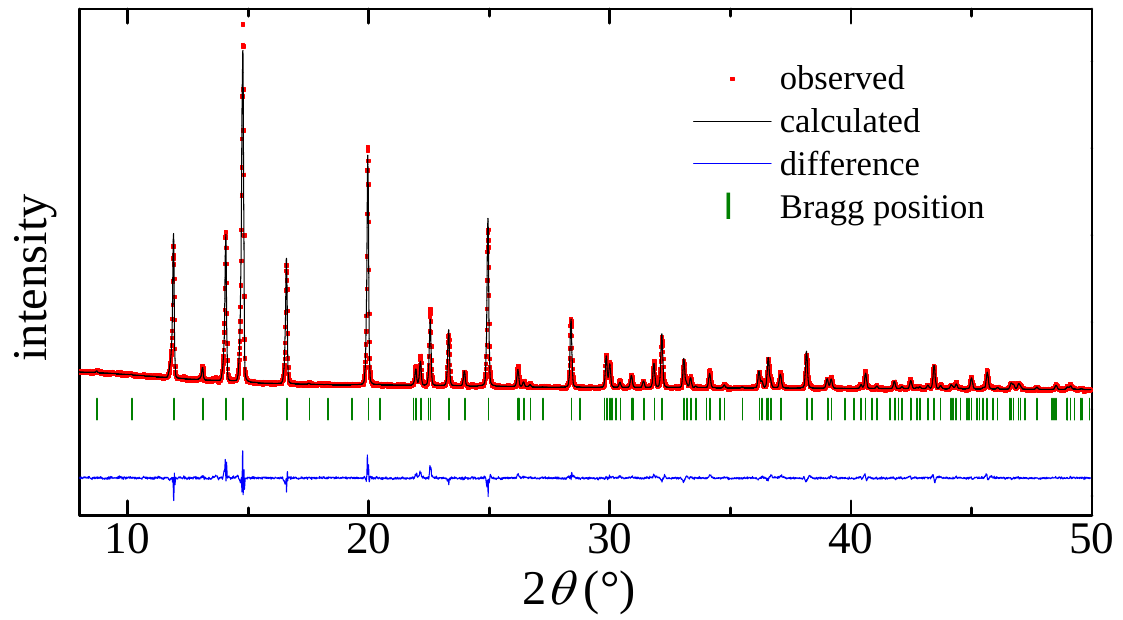}
	\caption{Room temperature powder X-ray diffraction spectrum and analysis on \LaOFeSe. Red symbols show the
  observed intensity. The black solid line represents the refined spectrum. Vertical green markers
  show the positions of the calculated Bragg reflections. The lower blue line is the difference between observed and
  calculated intensities.}
	\label{fig:XRD}
\end{figure} 

For $^{57}$Fe~M\"ossbauer spectroscopy a thin absorber of 2.6~mg/cm$^2$\, Fe area density was prepared by evaporation of a methanol suspension
of the powdered sample. 
A Wissel spectrometer was operated in sinusoidal mode and 
photons were detected with a KeTek-SDD detector.

$^{139}$La~NMR experiments in a temperature range of $5.0$~K~$\leq~T~\leq~170$~K were performed by
conventional pulsed NMR techniques with the duration of a $\pi/2$-pulse equal to 6~$\mu$s. 
The NMR data were recorded in a 8~T superconducting magnet system with a $^4$He variable temperature insert, 
using a Tecmag LapNMR spectrometer.
The polycrystalline powder was placed in a glass tube inside a Cu coil with a frequency of the resonant circuits of 44.0~MHz and 25.5~MHz.  
The spin-lattice relaxation rate was measured by using the saturation recovery method. 

An extensive search was carried out to resolve 
the $^{77}$Se signal, however without success.
Since the crystallographic Se site is located much closer to the Fe than the La site, a strong magnetic hyperfine coupling can lead to
a nuclear relaxation too fast to be measured.    

Zero magnetic field $\mu$SR spectra were recorded using the GPS instrument at the PSI Villigen, Switzerland, in
a temperature range $5~\le~T~\le~160$~K. The data were analyzed using the free software package \textsc{MUSRFIT}.\cite{Suter-PhysProcedia-2012} 

\section{$^{57}$Fe M\"ossbauer spectroscopy}
\label{sec:MoeSR}
 $^{57}$Fe M\"ossbauer spectroscopy allows to study microscopically the strength and alignment of ordered Fe magnetic moments 
 with respect to the electric field gradient (EFG) principal axis within the distorted
 FeO$_2$Se$_4$ octahedra. Moreover it is sensitive to temperature dependent changes of the electronic surrounding via the electric quadrupole
 interaction. 
 
\begin{figure}
	\centering
	\includegraphics{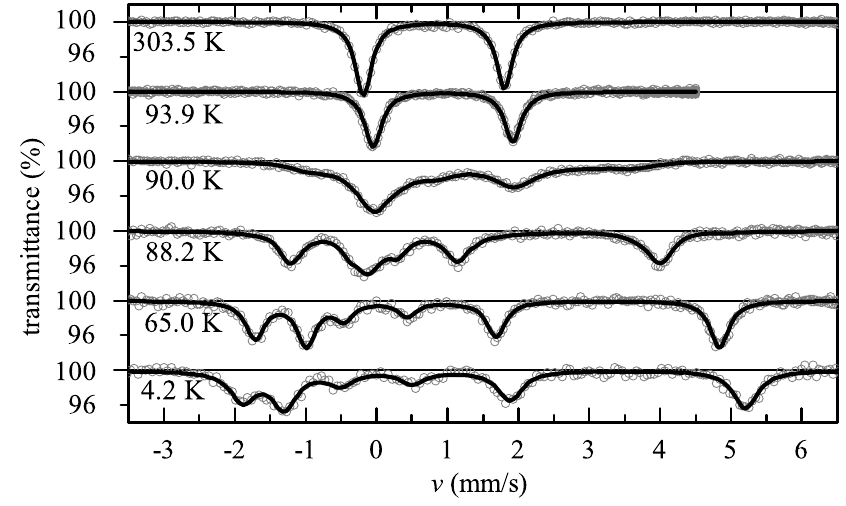}
	\caption{Temperature dependence of $^{57}$Fe M\"ossbauer spectra (open circles). Solid lines represent fits as described in the text. The paramagnetic spectra consist of an
  asymmetric doublet due to texture effects. A single sextet indicates full long range magnetic order rapidly arising below $T_\mathrm{N}$.}
  \label{fig:MoessbauerSpectra}  
\end{figure}

\begin{figure}
	\centering
	\includegraphics{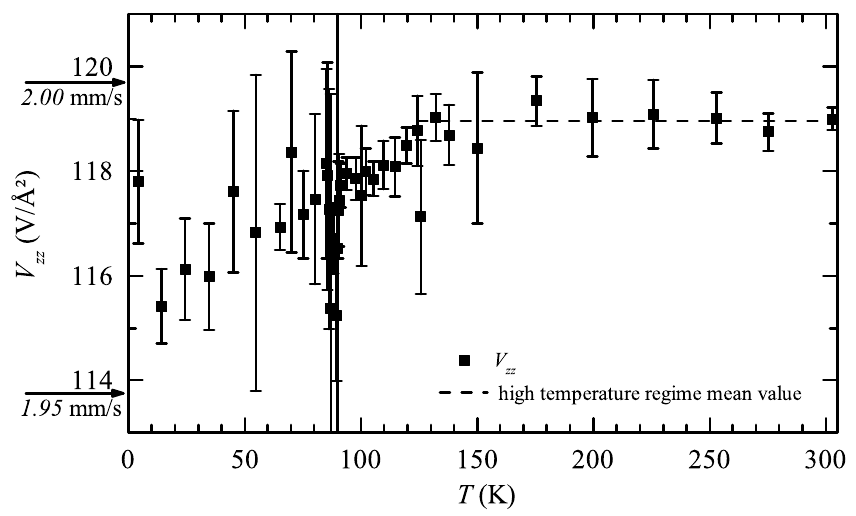}
	\caption{Temperature dependence of the principal axis component 
  $V_{zz}$ of the EFG. The dashed line indicates the mean value for $T~>~120$~K.
  The anomaly of $V_{zz}(T)$ occurs at the same temperature as that of $c$-axis thermal expansion. \cite{Free-2010-PRB}}
	\label{fig:FieldGradient}
\end{figure}

 Figure \ref{fig:MoessbauerSpectra} shows the $^{57}$Fe M\"ossbauer spectra at selected temperatures.
  In the paramagnetic regime 
 down to $T=120$~K we find a single asymmetric doublet with a quadrupole splitting of $\Delta v_{\mathrm{QS}}=1.98$~mm/s 
 (Fig. \ref{fig:FieldGradient}). 
 This corresponds to a principal EFG component of $V_{zz}=119~\mathrm{V}/\mathring{\mathrm{A}}^2$ 
  and is close to the values found in earlier studies.\cite{Fuwa-JPCM-2010,Lei-PRB-2012} 
The asymmetric absorption intensity of the doublet is interpreted in terms of texture caused by the flaky shape of the crystallites.  
These flakes are suspected to be oriented flat on the bottom of the sample holder,
resulting in a preferred orientation of the $c$-axis parallel to the $\gamma$-beam. With decreasing temperature
below $T_\mathrm{N}$ a sextet is developed, indicating long range magnetic ordering.
Due to the $D_{2h}$ symmetry of the $^{57}$Fe site an asymmetric 
electric field gradient, with asymmetry $\eta>0$, is expected. 
The analysis of the spectra is done by diagonalization of the full static hyperfine hamiltonian
\begin{eqnarray}
    H_\mathrm{s} & = & {\frac{eQ V_{zz}}{4I(2I-1)}}\left[(3I_z^2-I^2)+\frac{\eta}{2}(I_+^2+I_-^2)\right]  \\
    &&       -g_I\mu_\mathrm{N}B_\mathrm{hyp}\left(\frac{I_+e^{-i\phi}+I_-e^{+i\phi}}{2}\sin\theta+I_z\cos\theta\right)\nonumber ,
\label{FSHamiltonianMoes}
\end{eqnarray}
with nuclear spin operators $I_z$, $I_+~=~I_x+iI_y$, and $I_-~=~I_x-iI_y$.
 $B_\mathrm{hyp}$ is the hyperfine field at the $^{57}$Fe site. $Q$, $g_I$, and $\mu_\mathrm{N}$ denote the nuclear quadrupole moment, 
 g-factor, and magneton. The polar angle $\phi$ and the azimuthal angle $\theta$ describe the orientation of
the Fe hyperfine field $B_\mathrm{hyp}$ with respect to the EFG $z$-axis.
Due to the preferred orientation of the sample flakes parallel to the bottom of the sample holder,
i.e. the preferred orientation
of the $\gamma$-beam parallel to the $c$-axis, a texture function
\begin{equation}D(\phi_\gamma,\theta_\gamma)= \sin^N\phi_\gamma | \cos^N\theta_\gamma | = \cos^N \epsilon\end{equation}
is introduced to account for corrections of the line intensities with respect to the geometrical setup.    
Here the polar angle $\phi_\gamma$ and the azimuthal angle $\theta_\gamma$ describe the orientation of
the $\gamma$-beam with respect to the EFG $z$-axis. Additionally the equivalent description 
by the angle $\epsilon$ between the crystallographic $c$-axis and the $\gamma$-beam is given.
With this the preferred orientation can be smoothly tuned by the model parameter $N$.

In the AFM ordered state we find constant $N~=~1.14$ and $\eta~=~0.15(3)$ within error bars.
While a simple point charge model calculation estimates $\eta~<~0.146$ in \LaOFeSe, a smaller value was reported for \NdOFeSe 
 ($\eta~\approx~0.1$). \cite{Fuwa-JPCM-2010}
\begin{figure}
	\centering
	\includegraphics{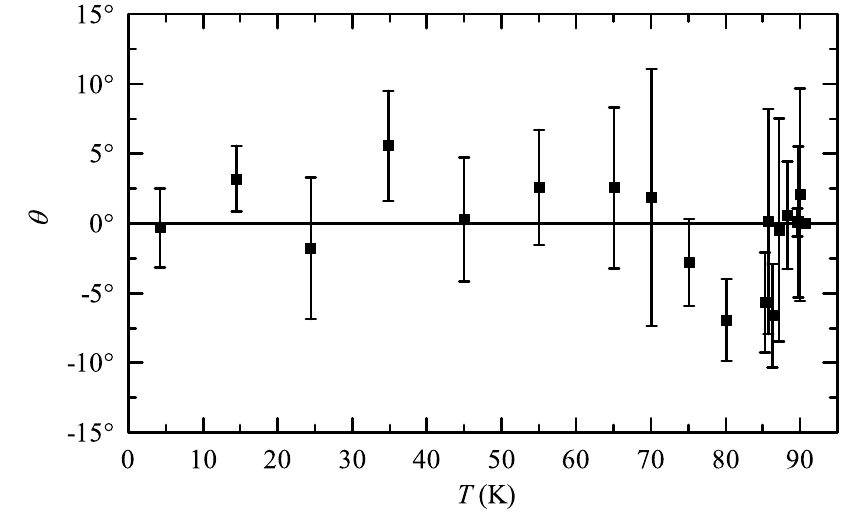}
	\caption{The angle $\theta$ between the principal EFG $z$-axis  and the $^{57}$Fe hyperfine field $\mathbf{B}_\mathrm{hyp}$ 
  in the AFM ordered state obtained from a fit to M\"ossbauer spectra.}
	\label{fig:theta}
\end{figure}
\begin{figure}
	\centering
	\includegraphics{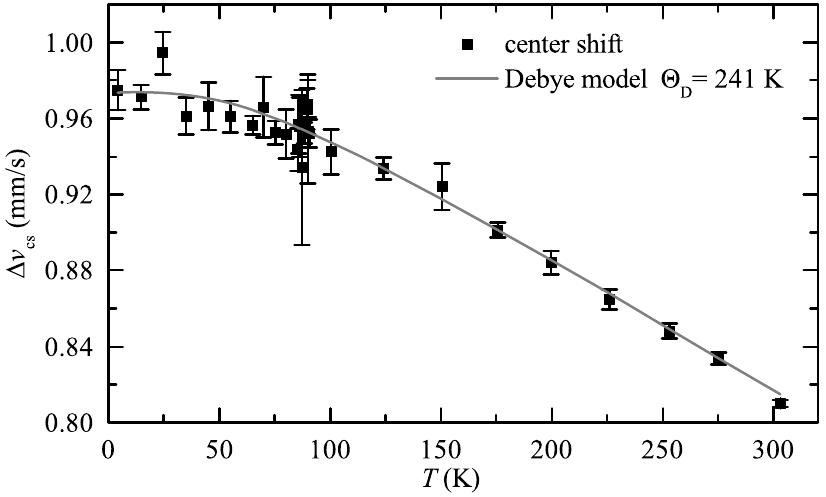}
	\caption{Temperature dependence of the center shift $\Delta v_\mathrm{cs}$.}
	\label{fig:CenterShift}
\end{figure}
\begin{table} 
\begin{tabular}{c r@{.}l r@{.}l r@{.}l}
\hline \hline
                                   & \multicolumn{2}{c}{$T_\mathrm{N}$(K)} & \multicolumn{2}{c}{$B_\mathrm{sat}$(T)} & \multicolumn{2}{c}{$\beta$}    \\
\hline
\LaOFeSe $^\mathrm{*}$             & 90&1                                  &  20&3                                   & 0&133                          \\
\LaOFeSe \cite{Fuwa-JPCM-2010}     & \multicolumn{2}{c}{-}                 &  \multicolumn{2}{l}{21}                 & \multicolumn{2}{c}{-}          \\
\NdOFeSe \cite{Fuwa-JPCM-2010}     & 88&3                                  &  \multicolumn{2}{l}{20}                 & 0&104                          \\
\SrFFeS  \cite{Kabbour-JACS-2008}  & 106&2                                 &  20&65                                  & 0&15                           \\
\BaFFeSe \cite{Kabbour-JACS-2008}  & 83&6                                  &  20&4                                   & 0&118                          \\
\hline
\end{tabular}
	\caption{Comparison of $^{57}$Fe M\"ossbauer studies on different oxychalcogenide systems with $\emph{Tm}~=~\mathrm{Fe}$. 
  The ordering temperature $T_\mathrm{N}$, 
  low temperature $^{57}$Fe saturation hyperfine field $B_\mathrm{sat}$ and the critical exponent $\beta$ are found to vary only in small 
  ranges. * denotes results of the present work.} 
\label{table:MoessbauerTable} 
\end{table}
 $N$ and $\eta$ are fixed for further analysis of $V_{zz}$ and $B_\mathrm{hyp}$.
While decreasing the temperature the magnetic volume fraction increases rapidly. At 88~K the sample
 is 100\% magnetic. 
At the lowest measured temperature $T~=~4.2$~K the spectrum is described by a Zeeman splitting with maximum 
hyperfine field $B_\mathrm{sat}~=~20.3(1)$~T. 
All spectra are fitted with a single Fe site. 
The isomer shift of 0.82~mm/s at room temperature is typical for a high spin state of Fe$^{2+}$. \cite{Shirane-PR-1962}
The absolute value of the $^{57}$Fe hyperfine field at low temperatures $B_\mathrm{sat}$
is close to that of other oxychalcogenide systems (Tab.~\ref{table:MoessbauerTable}).
For \SrFFeS a saturation field $B_\mathrm{sat}~=~20.65$~T and the low temperature ordered 
Fe magnetic moment $m_\mathrm{Fe}=3.3(1)~\mu_\mathrm{B}$ is reported.\cite{Kabbour-JACS-2008,Zhao-PRB-2013}
Taking into account the very similar local coordination of the Fe ions in \OX and \SrFFeS, 
the same conversion factor $A=B_\mathrm{sat}/m_\mathrm{Fe}=6.26~\mathrm{T}/\mu_\mathrm{B}$ between 
the $^{57}$Fe hyperfine field
and the ordered magnetic moment, as calculated from the \SrFFeS values, is valid for both materials. 
From this a low temperature ordered Fe magnetic moment 
of $m_\mathrm{Fe}= B_\mathrm{sat} / A=3.2~\mu_\mathrm{B}$ in \OX is deduced from
the saturation field.

M\"ossbauer spectroscopy clearly shows that $\theta~=~0$ in the magnetically ordered regime, 
i.e. the magnetic hyperfine field $\mathbf{B}_\mathrm{hyp}$ is oriented parallel 
to the $z$-axis of the EFG principal axis system (Fig.~\ref{fig:theta}).
Considering that the EFG strongest component is aligned along the O-Fe-O chains (also calculated in a LSDA+U approach), \cite{Fuwa-JPCM-2010}
this proves that the ordered Fe magnetic moments are oriented parallel to O-Fe-O chains, resulting in a non-collinear magnetic order. 
This was already concluded for the Nd-system. \cite{Fuwa-JPCM-2010} Considering  AFM $J_\mathrm{nnn1}$ intraplane exchange interactions, 
three different models of non-collinear order are possible (see Fig.~\ref{fig:Structure}).

Fitting the sublattice magnetization for temperatures above $0.6~T_\mathrm{N}$,
 $M(T)~\propto~B_\mathrm{hyp}(T)~\propto~(1-T/T_\mathrm{N})^\beta$ yields a transition 
 temperature $T_\mathrm{N}~=90.1~\pm~0.1~$K.
The critical exponent $\beta~=~0.133$ is close
to that of the magnetic square planar Ising model ($\beta~=~1/8$).

\begin{figure}
	\centering
		\includegraphics{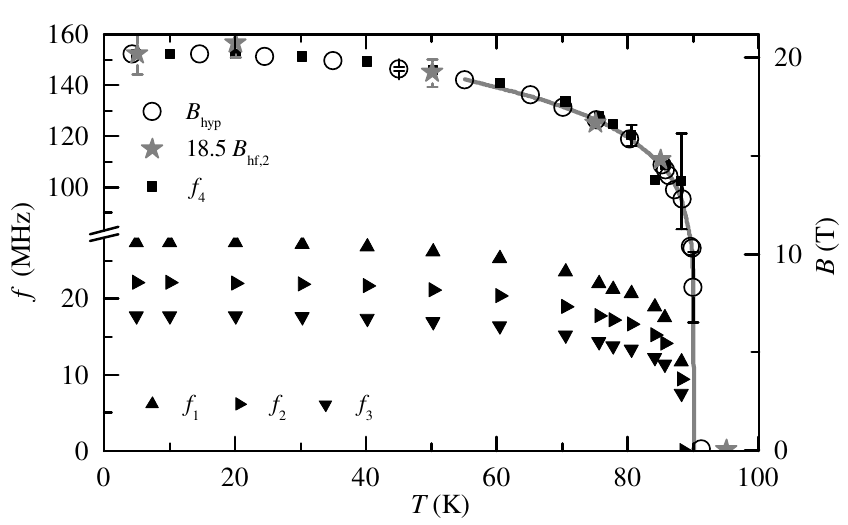}
	\caption{Magnetic order parameter determined from M\"ossbauer, NMR, and $\mu$SR measurements. 
  Open circles represent $^{57}$Fe hyperfine fields fitted to M\"ossbauer spectra.
  Gray stars refer to $^{139}$La hyperfine fields $B_\mathrm{2,\perp}$ scaled by a factor of 18.5.
  Black solid symbols show fitted $\mu$SR frequencies $f_1 - f_4$ as described in the text.
  Representative error bars are shown.
  The gray line shows the fit of sublattice magnetization of the M\"ossbauer data according to the text.}
	\label{fig:orderParameter}
\end{figure}

Figure~\ref{fig:FieldGradient} shows the temperature dependence of the EFG component $V_{zz}(T)$.
$V_{zz}(T)$ is constant at high temperatures above 120~K and then decreases gradually below.
This onset temperature is in accordance with an anomaly of the thermal expansion along the $c$-axis, detected in neutron scattering experiments.\cite{Free-2010-PRB}
The Debye-Waller-factor as well as the quadratic Doppler shift can be described in the Debye model (Fig.~\ref{fig:CenterShift}). 
For both datasets a Debye temperature $\Theta_\mathrm{D}~=~241$~K is obtained and no anomaly
is found below 120 K. 

\section{$^{139}$La NMR}
\label{sec:NMR}
\begin{figure}
\includegraphics{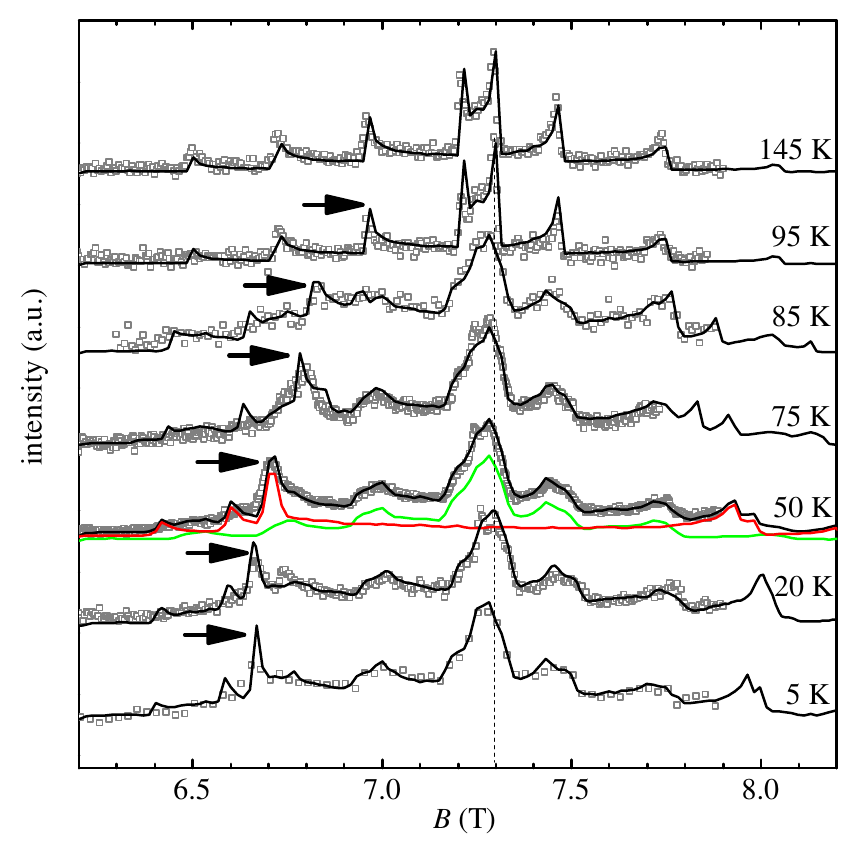}
\caption{\label{fig:NMR} Temperature dependence of the $^{139}$La field-swept NMR spectra at
 44.0~MHz (20~K spectrum at 44.1~MHz). For $T~>~T_\mathrm{N}$ typical non-magnetic $I~=~7/2$ powder spectra are observed. 
 Black arrows indicate the position of $^{139}$La signal in the ordered regime stemming from a satellite position at high temperatures.
 Black solid lines are simulated spectra.
 Green and red solid lines at 50~K represent calculated sub spectra due to magnetically non-equivalent sites (see text). 
 The vertical dash-dotted line indicates 
 the field where $^{139}(1/T_1)$ were measured. 
} 
\end{figure}

In this section we present field-swept $^{139}$La NMR spectra and spin-lattice relaxation data. Figure \ref{fig:NMR}
shows $^{139}$La spectra at different temperatures. 
\begin{figure}
\includegraphics{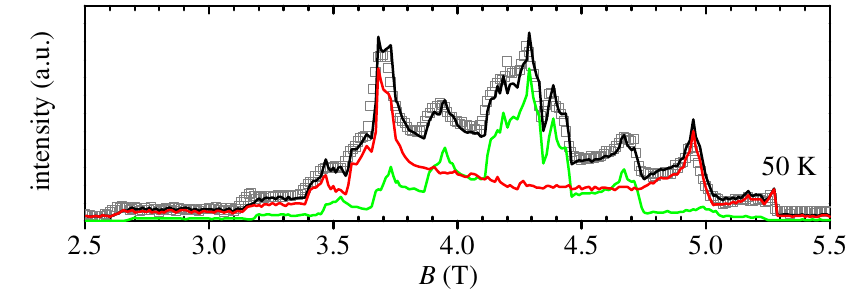}
\caption{\label{fig:NMR25} Gray squares show $^{139}$La field-swept NMR spectrum at
 25.5~MHz and 50~K. Green and red solid lines represent calculated spectra due to the two magnetically non-equivalent sites (see text).
 The total simulated spectrum is shown as black line.     
} 
\end{figure}
At high temperatures a typical nuclear spin $I=7/2$ powder line shape is observed. For $T~>~T_\mathrm{N}$ the spectra consist of 
a splitted central transition and three pairs of satellites. Note that the most outer satellite at the high field side 
was not observed at 44.0~MHz, but has been verified in measurements at a frequency of 25.5~MHz. 
In the paramagnetic state the deduced quadrupole frequency is $\nu=3.1$~MHz and the asymmetry parameter $\eta=0$. 
Due to the local $C_{4v}$ symmetry of La, a vanishing $\eta$ is reasonable and the EFG principal $z$-axis is assumed to match
the crystallographic $c$-axis. 
These values are fixed for the analysis of the magnetically ordered regime $T~<~T_\mathrm{N}$, where two effects lead to more complex spectra:
1. A significant broadening of the central transition and the satellites is 
attributed to static hyperfine fields.
2. Additional $^{139}$La intensity appears at 85~K, near 6.8~T and strongly shifts to lower fields, when cooling the sample.    
Since there is only one crystallographic La position in the unit cell, the NMR experiments show
two magnetically non-equivalent sites.

To analyze the field-swept spectra, the nuclear Hamiltonian
\begin{equation}
H= - \hbar \gamma (\mathbf{B}+\mathbf{B}_\parallel + \mathbf{B}_\perp) \mathbf{I} + h \frac{\nu}{6} [3 I_z^2 - \mathbf{I}^2 + \eta (I_x^2 - I_y^2) ]
\label{equ:NMRHamiltonian}
\end{equation}
was diagonalized for
25000 random orientations of the external field $\mathbf{B}$, with 
respect to the EFG principal $z$-axis, at each sweep step ($\gamma/2\pi~=6.0146~\mathrm{MHz}/\mathrm{T}$).
 $\mathbf{B}_\parallel$ and $\mathbf{B}_\perp$ are hyperfine fields parallel and perpendicular to the $a$-$b$ plane.  
Transitions of nuclear sublevels
corresponding to 44.0~MHz (in a window of 70~kHz corresponding to the signals FWHM) provide the intensity
for a given field point.

While the modeled spectra yield an accurate description of the experimental data without any additional 
internal field ($\mathbf{B}_\parallel=0$~mT and $\mathbf{B}_\perp=0$~mT) at temperatures $T~>~T_\mathrm{N}$,
two local fields $\mathbf{B}_\mathrm{1}$ and $\mathbf{B}_\mathrm{2}$ at magnetically non-equivalent $^{139}$La sites must be 
introduced to describe the observed spectra in the ordered state.
  
\emph{Site 1:} Choosing $\mathbf{B}_\mathrm{1}$ parallel to the $a$-$b$ plane
results in a broadening of the simulated spectra. The green lines in Fig. \ref{fig:NMR} and \ref{fig:NMR25} are simulations
with $|\mathbf{B}_\mathrm{1}|~=~B_{1,\parallel}~=~$35~mT and reproduce some features of NMR line broadening at $T~=~$50~K.
Any additional component parallel to the $c$-axis will impair the simulation and we find $B_{1,\perp}~=~$0~mT. 

\emph{Site 2:} To describe the second signal, a simulation with $\mathbf{B}_\mathrm{2}~\parallel~\mathbf{c}$ is 
shown as red line. 
Here ${B}_\mathrm{2,\perp}=1040$~mT at 50~K. 
Additional components of the hyperfine field parallel to the $a$-$b$ plane, in the order of $35$~mT,
will not interfere the simulation.
 
Black lines are spectral sums of equal weighting for both $^{139}$La sites and
model our field-swept data very well in the magnetically ordered regime. 
The temperature dependency of $B_\mathrm{2,\perp}$ is shown in Fig. \ref{fig:orderParameter}.     
Additional field-swept spectra at a resonance frequency of 25.5~MHz yields similar 
results and confirm the magnetic origin of line broadening and shifting (Fig.~\ref{fig:NMR25}).

\begin{figure}
\includegraphics[width=0.45\textwidth]{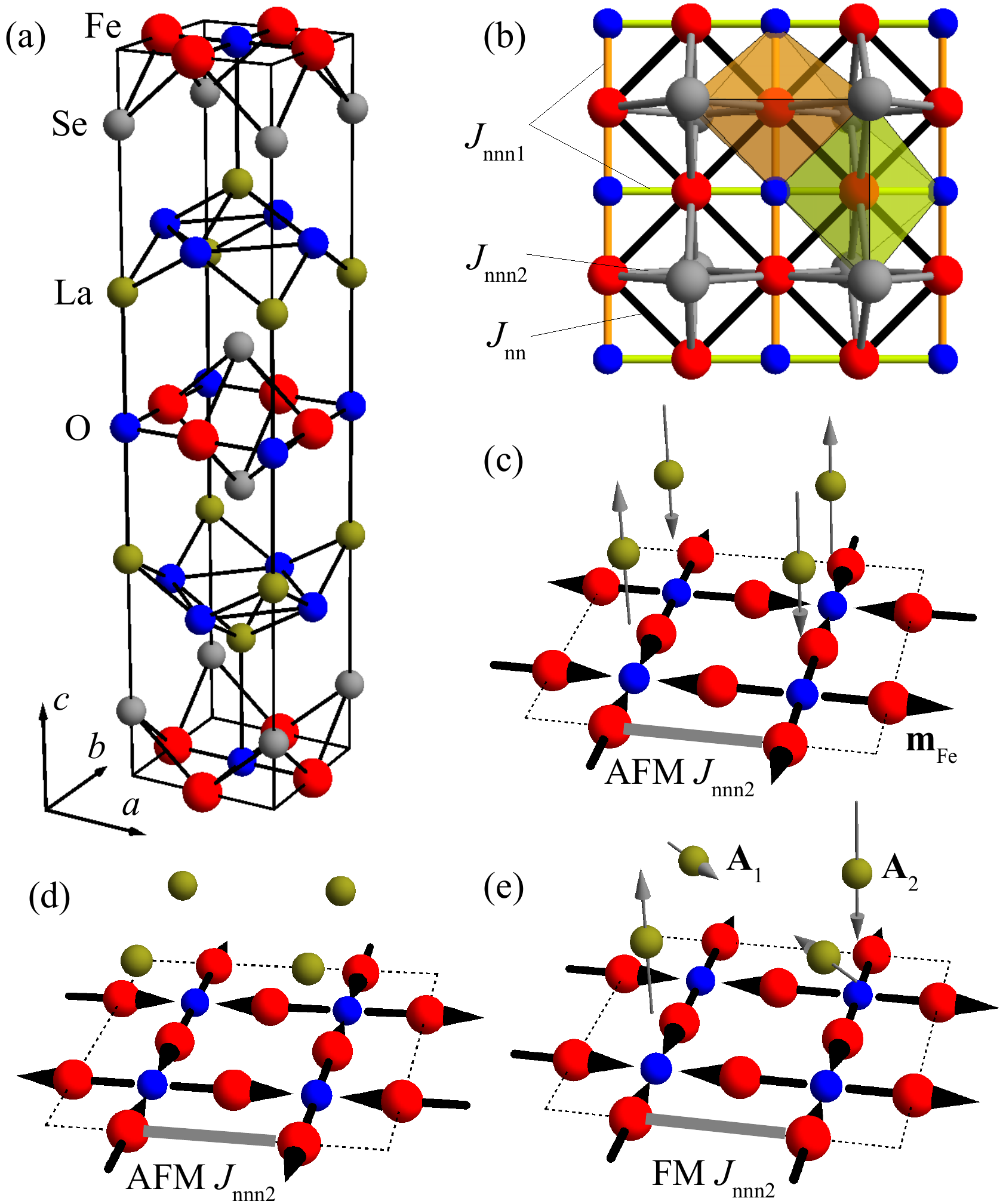}
 \caption{\label{fig:Structure}(a) Crystallographic unit cell of \LaOFeSe. \\
 (b) Magnetic exchange interactions in the $\mathrm{Fe}_2\mathrm{O}\mathrm{Se}_2$ layer.
 Green and orange distorted octahedra sketch the coordination of Fe ions along perpendicular Fe-O-Fe chains.
 Green and orange lines symbolize the $J_\mathrm{nnn1}$ super exchange pathway along these chains.
 Gray and black lines represent $J_\mathrm{nnn2}$ and $J_\mathrm{nn}$.    
 For AFM $J_\mathrm{nnn2}$ and moments aligned along O-Fe-O chains there are three realizations of magnetic 
 structure as discussed in the literature.\\
 (c)-(e) Sketches of the $\mathrm{Fe}_2\mathrm{O}$ plane for the three different non-collinear models of magnetic structure.
 Additionally representative La sites are shown. The gray arrows
 indicate the calculated dipolar hyperfine coupling.\\ 
 (c) AFM $J_\mathrm{nnn2}$ yields strong hyperfine fields at all $^{139}\mathrm{La}$ sites. \\
(d) AFM $J_\mathrm{nnn2}$ yields no hyperfine fields at all $^{139}\mathrm{La}$ sites. \\
(e) Ferromagnetic $J_\mathrm{nnn2}$ interaction results in the model of magnetic order as described for \SrFFeS and yields 
  magnetically non-equivalent $^{139}$La sites.}
\end{figure}

We calculated the hyperfine coupling constants at the $^{139}$La sites, using localized Fe magnetic dipole moments 
for three non-collinear models of magnetic order (Fig.~\ref{fig:Structure}).
For each model (c) and (d), 
equivalent values of hyperfine coupling parallel and perpendicular to the $c$-axis
are obtained at all $^{139}$La sites. 
Therefore those models cannot describe our NMR field swept spectra in the magnetically ordered state.
Only model (e) results in magnetically non-equivalent $^{139}$La sites:

\emph{Site 1:} For site 1 the small hyperfine coupling is parallel to the $a$-$b$ plane.
The value of the components reads $A_{1,\parallel}=8.3~$mT$/\mu_\mathrm{B}$ and $A_{1,\perp}=0~$mT$/\mu_\mathrm{B}$.
The parallel component primarily arises due to the nnn FeO$_2$ plane,
while all contributions from the nn plane cancel with respect to the symmetry.

\emph{Site 2:} For site 2 a strong hyperfine coupling is essentially parallel to the $c$-axis.
The value of the components is $A_{2,\parallel}=8.3~$mT$/\mu_\mathrm{B}$ and $A_{2,\perp}=81.5~$mT$/\mu_\mathrm{B}$.
The strong component parallel to the $c$-axis primarily arises due to the coupling to
Fe magnetic dipolar moments located in the nn FeO$_2$ plane. 
Again the parallel component is dominantly due to the nnn FeO$_2$ plane. 

This model is identified to be consistent with our experimental results.

Comparing the hyperfine fields deduced from the NMR field-swept experiment to the ordered magnetic Fe moment deduced
from M\"ossbauer spectroscopy, we obtain a hyperfine coupling parallel to the $a$-$b$ plane 
of $A_{1,\parallel}^\prime~=~11.4~\mathrm{mT}/\mu_\mathrm{B}$ for site 1 and
$A_{2,\perp}^\prime~=~340~\mathrm{mT}/\mu_\mathrm{B}$ for site 2 .
The dipolar model yields a good qualitative description of our field-swept spectra for both $^{139}\mathrm{La}$ sites.
For site 1 also a good quantitative agreement is found (37\% difference between experimental and calculated 
hyperfine coupling). For site 2 the experimental value $A_{2,\perp}^\prime$ is approximately four times larger than the 
calculated $A_{2,\perp}$.
This shows a limitation of the dipole approximation which does not take into account additional 
contributions of transfered hyperfine coupling.

\begin{figure}
\includegraphics[width=0.45\textwidth]{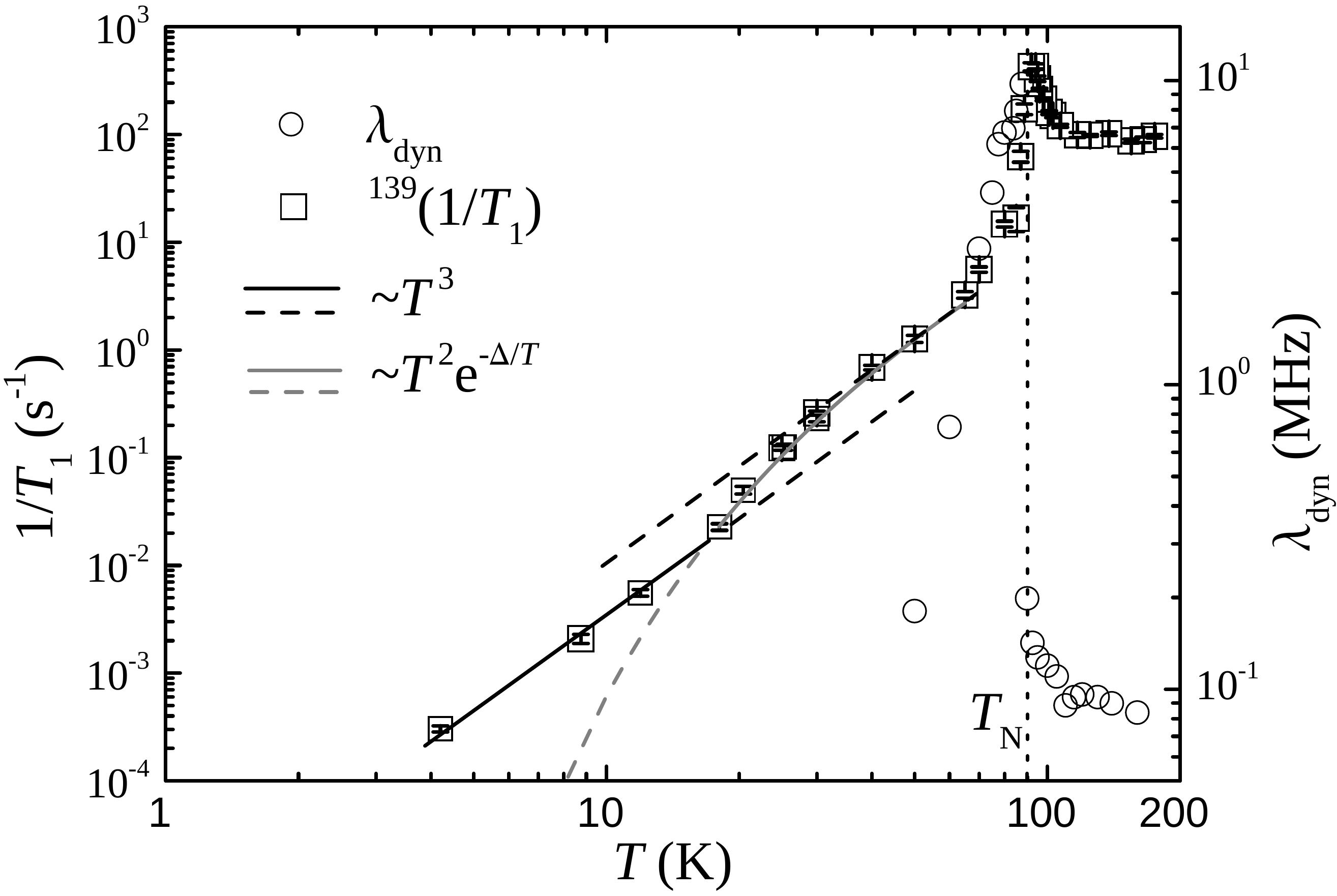}
\caption{The temperature dependence of $^{139}(1 / T_1)$ is shown as open squares.
The upper black dashed line shows the asymptotic $T^3$ power law for $T>>\Delta$.
The gray solid line and gray dashed line represents the model of activated two-magnon processes.
The lower black lines show a $T^3$ power law behavior which is hidden for $T>20$~K. 
Open circles show dynamic relaxation rates $\lambda_\mathrm{dyn}$ 
of the muon-spin polarization.}
\label{fig:4}
\end{figure}
The $^{139}$La spin-lattice relaxation rate $^{139}(1/T_1)$ was measured by conventional
saturation recovery method, exciting the central transition line at 44.0~MHz, and 7.297~T (marked by the vertical dash-dotted line 
in Fig.~\ref{fig:NMR}). 
Recovery curves are fitted by
\begin{eqnarray}
1-\frac{M(t)}{M(0)}&=&\frac{1}{84}\exp(- t/T_1) + \frac{3}{44}\exp(-6 t/T_1) \nonumber\\ 
&& +\frac{75}{364}\exp(-15 t/T_1) + \frac{1225}{1716}\exp(-28 t/T_1) \nonumber ,
\end{eqnarray}
taking into account magnetic transition probabilities $1/T_1$ for a relaxation of a spin $I=7/2$ nuclei central transition.
$M(t)$ is the value of the nuclear magnetization at a time $t$ 
after the saturation pulse and $M(0)$ the equilibrium magnetization.

Figure~\ref{fig:4} shows the $^{139}(1/T_1)$ vs $T$ plot together with the dynamic relaxation rate of the muon-spin-polarization. 
For $T>T_\mathrm{N}$, $^{139}(1/T_1)$ 
monitors the slowing down of magnetic fluctuations while approaching $T_\mathrm{N}$ with respect to equivalent $^{139}$La sites.
For $T<T_\mathrm{N}$ the experiments at 7.297~T mainly probe site 1 and
 the spin-lattice relaxation is strongly reduced within an interval of 5~K.
For $T<65$~K we find $^{139}(1/T_1)\propto T^3$ close to a power law.
In the ordered state of AFM insulators the relaxation of nuclear spins in the presence of a gap $k_B\Delta$  
in the magnetic excitation spectrum is mainly driven by Raman processes.\cite{Moriya-PTP-1956}
For $T>>\Delta$ a $T^3$ behavior of the spin-lattice relaxation rate is expected for a dominating two-magnon 
process, while a three-magnon process will result in a $T^5$ power law.\cite{Beeman-PhysRev-1968}
The deviations of $^{139}(1/T_1)$ from the $T^3$ law towards lower temperatures indicates
the crossover to the regime $T<<\Delta$, where the thermal activation of gapped excitations will lead to 
$^{139}(1/T_1) \propto T^2~e^{-\Delta/T}$. The gray line in Fig.~\ref{fig:4} refers to $\Delta=55$ K, which is
close to the observed gap in the reference.\cite{McCabe-PRBR-2014} 
In the low temperature regime the relaxation due to the gapped excitations is
rapidly suppressed. For $T<20$~K it is dominated by a process which again reveals a
 power law $^{139}(1/T_1)\propto T^3$. 
\section{Zero field $\mu$SR experiments}
\label{sec:muSR}
\begin{figure}
	\centering
		\includegraphics[width=0.47\textwidth]{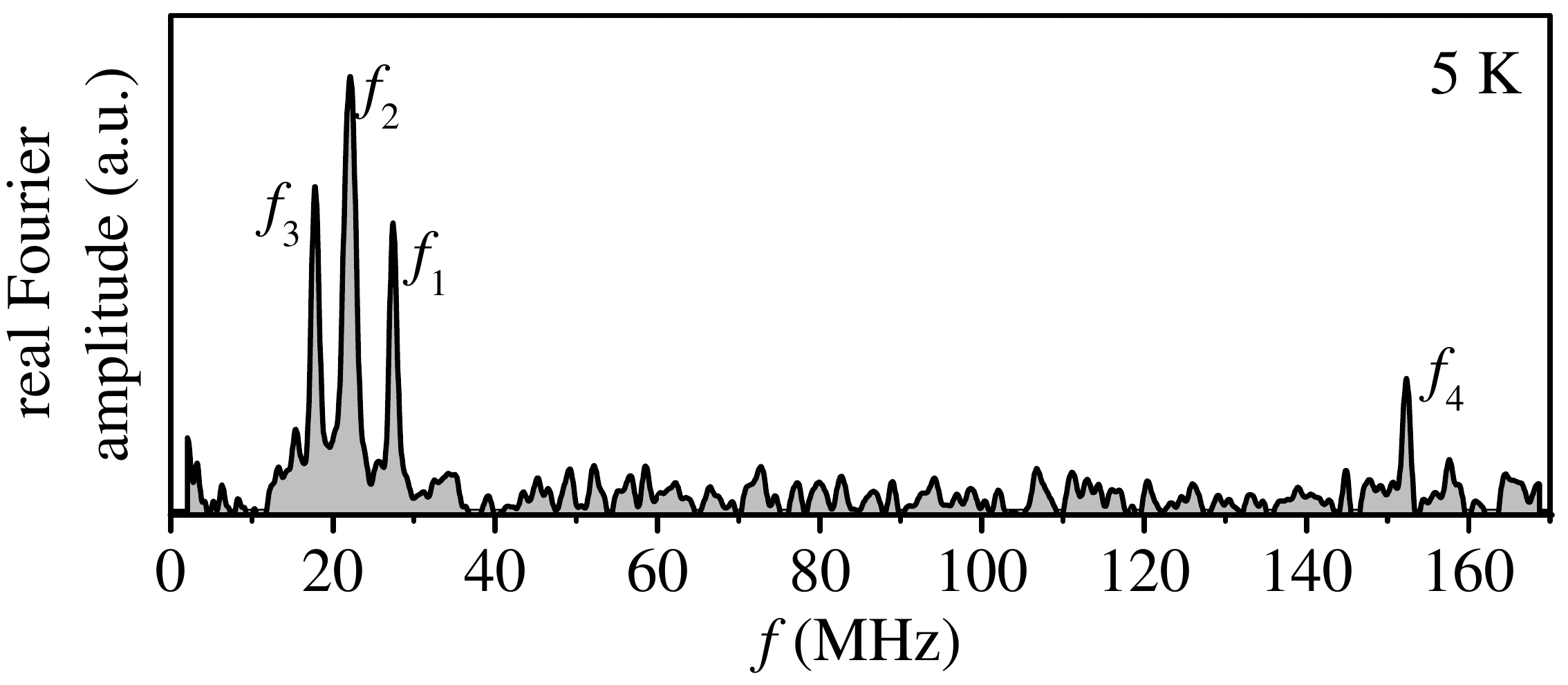}
	\caption{The real Fourier amplitude of the muon spin polarization measured at $T=5$~K 
  shows four well defined muon spin precession frequencies.  
  }
	\label{fig:RFA}
\end{figure}
\begin{figure}
	\centering
		\includegraphics[width=0.45\textwidth]{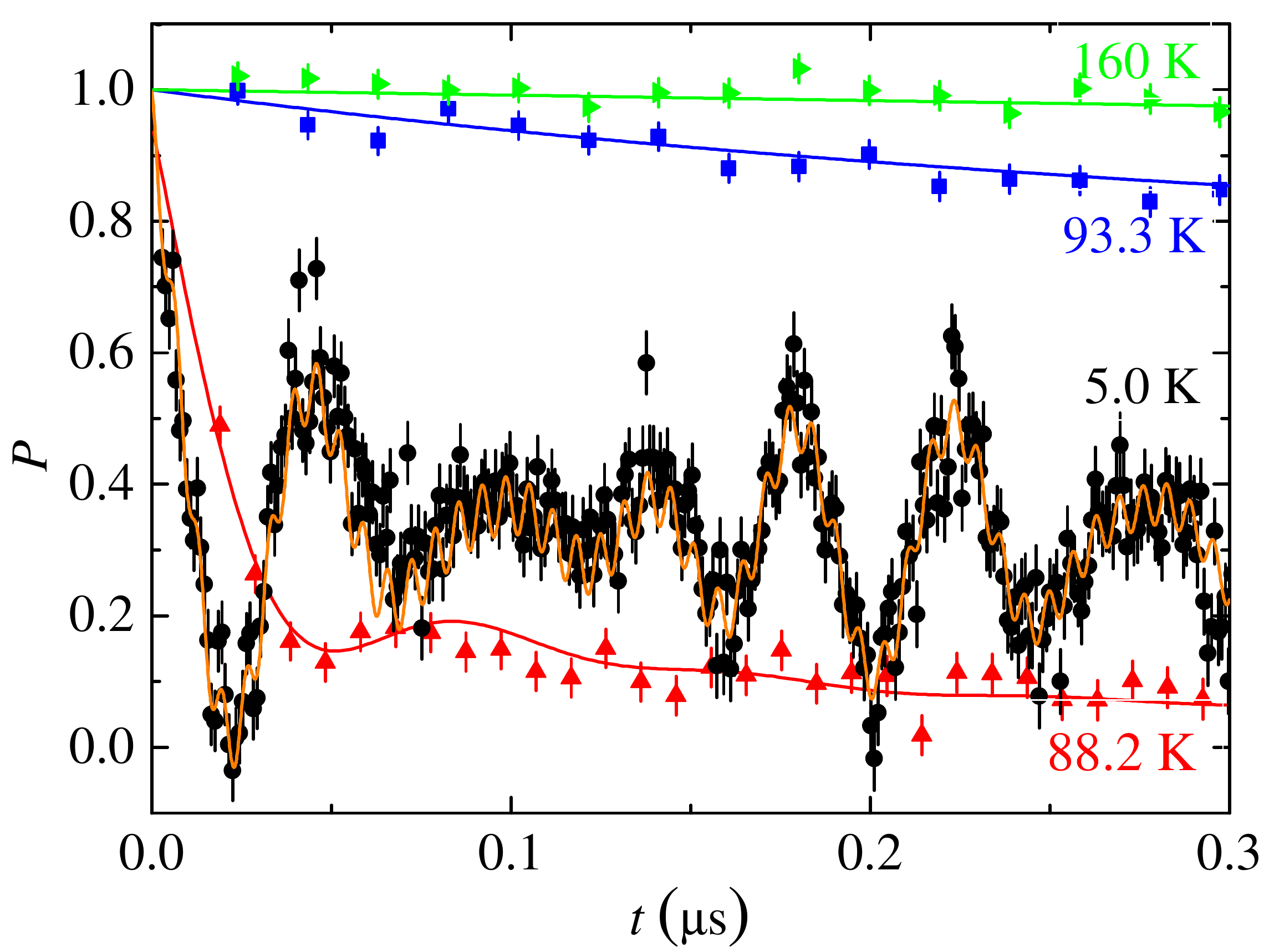}
	\caption{Representative zero field $\mu$SR spectra and fits of $P(t)$ according 
  to Eq.~\ref{equ:muSRFitHT}~and~\ref{equ:muSRFitLT}.
 $T=93.3$~K and $T=160$~K: In the absence of long range magnetic order $P(t)$ relaxes purely exponential.
    $T=5$~K: $P(t)$ is fitted with a superposition of oscillating and non-oscillating fractions.
   $T=88.2$~K: Oscillating and non-oscillating fractions are strongly affected by dynamic relaxation of the polarization.  
  }
	\label{fig:ZF}
\end{figure}
In zero field $\mu$SR experiments the time dependent polarization $P(t)$ of an initially polarized muon-spin ensemble is obtained. 
After implantation, the positive muons thermalize very fast and come to rest at interstitial lattice sites.
The muon decays into two neutrinos and a positron which is detected. Because the direction of the positron emission is favored
along the direction of the muon spin at the moment of the decay, an ensemble average of several million decays provides $P(t)$.
Representative spectra on powder samples of \OX at selected temperatures are shown in Fig.~\ref{fig:ZF}.

For $T~>~T_\mathrm{N}$ the polarization $P(t)$ is described by an exponential relaxation with two fractions $P_\mathrm{dyn}~+~P_\mathrm{dyn}^\prime~=~100\%$:\cite{muSR}  
\begin{equation}
	P(t)~=~P_\mathrm{dyn} e^{-\lambda_\mathrm{dyn}t} + P_\mathrm{dyn}^\prime e^{-\lambda_\mathrm{dyn}^\prime t}.
\label{equ:muSRFitHT}
\end{equation}
$\lambda_\mathrm{dyn}$ and $\lambda_\mathrm{dyn}^\prime$ denote relaxation rates of the time dependent muon spin polarization.
For $T~>~120$~K we find $P_\mathrm{dyn}~=~100\%$, indicating 
magnetically equivalent muon sites in the paramagnetic regime.

In accordance with the onset temperature of the $V_{zz}$ reduction and the corresponding anomaly of the thermal expansion we
identify a second fraction $P_\mathrm{dyn}^\prime~\approx~7\%$ below $T~=~120$~K. This fraction exhibits a rather
strong relaxation $\lambda_\mathrm{dyn}^\prime~>~1.7$~MHz~$>>~\lambda_\mathrm{dyn}$. We attribute this fraction of the $\mu$SR
signal to muons affected by a large hyperfine coupling constant which are therefore probably located at an interstitial site
close to a magnetic Fe moment. This conclusion will later be reinforced by the observation of an equivalent signal fraction
probing a strong static internal magnetic field in the low temperature, long range ordered magnetic phase (see below).  

 $\lambda_\mathrm{dyn}$ is shown in Fig.~\ref{fig:4} and monitors the slowing down 
of magnetic fluctuations while approaching $T_\mathrm{N}$ in the paramagnetic regime.

In the magnetically long range ordered regime $T~<~T_\mathrm{N}$ internal fields at the muon sites develop and therefore
spontaneous muon-spin precession is observed.
Taking into account the powder average, the polarization is described by 2/3 oscillating and 1/3 non-oscillating fractions.
A magnetic five site model for the oscillating fraction is identified from a Fourier transformation of the polarization $P(t)$
(Fig.~\ref{fig:RFA}),
which displays four well defined frequencies and a fast relaxing contribution.
The non-oscillating part of the polarization is described by an effective two site model, 
with 85\% dynamically depolarizing and 15\% non-depolarizing fractions:\cite{muSR}  
\begin{equation}
	P(t) =\frac{2}{3}\sum_{i=0}^4 P_i e^{-\lambda_{T_i}} \cos(2\pi f_i t) + \frac{1}{3}(\underbrace{0.85e^{-\lambda_\mathrm{dyn} t}}_{dyn.} + \underbrace{0.15}_{stat.} ).
\label{equ:muSRFitLT}
\end{equation}
Here $\lambda_{T_i}$ describe the depolarization of the oscillating signal fractions.
The fast depolarization of fraction $P_0$ with $\lambda_{T,0}~>~50$~MHz prohibits a determination of $f_0$. 
We find signal fractions $P_0~=~36.4\%,~P_1~=~12.2\%,~P_2=33.1\%,~P_3~=~10.6\%$, and $P_4~=~7.7\%$ at lowest measured temperature 
$T~=~5~\mathrm{K}$ and fix them for the analysis of the magnetically ordered regime. 
The obtained precession frequencies $f_1, f_2 , f_3$, 
and $f_4$ are shown in Fig.~\ref{fig:orderParameter}. In our analysis the ratios
of the frequencies $f_1, f_2$ and $f_3$ are determined from the 5~K spectrum and
fixed for the further analysis. The high frequency oscillation $f_4$ nicely scales
with the order parameter determined from $^{57}$Fe Mössbauer and $^{139}$La NMR
experiments within error bars. In correspondence to the presented analysis of the $^{57}$Fe Mössbauer order parameter
the frequencies $f_1, f_2$ and $f_3$ yield a consistent critical exponent $\beta=0.15(1)$ and transition
temperature $T_\mathrm{N}=88.5(8) K$.

Fraction $P_4$ is nearly equal to $P_\mathrm{dyn}^\prime$ and $f_4>>f_1,f_2,f_3$ indicates strong hyperfine coupling due to a close vicinity to 
magnetic Fe moments. We identify $P_4$ and $P_\mathrm{dyn}^\prime$ as stemming from crystallographically equivalent muon sites. 
Spontaneous precession frequencies due to internal magnetic fields probe the magnetic order parameter. 
We find $f_4$ proportional to $B_\mathrm{hyp}$ deduced
from M\"ossbauer spectroscopy and $B_\mathrm{2,\perp}$ deduced from $^{139}$La NMR experiments (Fig.~\ref{fig:orderParameter}).

We interpret the multiplicity of precession frequencies $f_0$-$f_3$ as due to magnetically non-equivalent sites in the ordered regime. 
Compared to the magnetic twofold splitting of
$^{139}$La spectra, a reduced symmetry of interstitial muon sites will lead to a more complex splitting of muon-spin precession frequencies. 

Below $T_\mathrm{N}$ the dynamic depolarization $P_\mathrm{dyn}$ is reduced strongly. 
Finally for $T~\le~50$~K only small residual depolarization $\lambda_\mathrm{dyn}~<~0.1$~MHz is fitted
(not shown in Fig.~\ref{fig:4}).
Here the relaxation of the different muon fractions successively drops below the sensitivity of the method
near 50~K.

\section{Summary}
\label{sec:Summary}
We have investigated magnetic order and spin dynamics in \OX by means of $^{57}$Fe M\"ossbauer spectroscopy, $^{139}$La NMR, and muon spin relaxation.
The analysis of the M\"ossbauer spectra puts the system close to related oxychalcogenide systems (Tab.~I). 
The spectra reveal the presence of magnetically equivalent Fe moments with non-collinear alignment parallel to O-Fe-O chains in the ordered state. 
This is in contrast to the collinear order model concluded in a neutron spectroscopy study but consistent with former theoretical predictions.
The transition temperature to magnetic order is $T_\mathrm{N}=90.1$~K. 
The critical exponent of the magnetic order parameter $\beta~=~0.133$ is close to that of the 2D Ising universality class.    

In NMR field-swept experiments we find two magnetically non-equivalent $^{139}$La sites in the magnetically ordered state.
This is shown to be consistent with only one model of non-collinear magnetic order discussed in the literature, 
i.e. the model shown in Fig.~\ref{fig:Structure}(d).  
This model was already suggested as one possible model for \NdOFeSe and was recently found to describe magnetic order in \SrFFeS. 

The observation of four muon-spin precession frequencies and a fast depolarizing signal in the ordered state in \OX is interpreted in terms of 
 two different muon sites and the complex model of magnetic order. 

Dynamic depolarization of the muon-spin ensemble and $^{139}$La spin-lattice relaxation 
reveal a slowing down of magnetic fluctuations while approaching the long range ordered state.
For $T<T_\mathrm{N}$ magnetic fluctuations are strongly suppressed. 
In the temperature range $20~\mathrm{K}<T<65~\mathrm{K}$ $^{139}(1/T_1)$ data probes
the cross over to the gapped regime $T<T_\Delta\approx 55~\mathrm{K}$, in consistence 
with inelastic neutron scattering data. \cite{McCabe-PRBR-2014}   
For $T<20~\mathrm{K}$ the strong reduction of spin-lattice relaxation rate
due to the excitation gap is finally dominated by another source of relaxation, again leading to
a power law behavior $^{139}(1/T_1)\propto T^{3}$. This is indicative for the presence
of low energy magnetic excitations below the gap energy.
In the magnetically ordered state of \OX a small residual dynamic depolarization 
hides the signature of the spin excitation band. \cite{McCabe-PRBR-2014}

Due to the peculiar local in-plane coordination auf Fe atoms by oxygen, the magnetism
in \OX is more anisotropic then in Fe pnictides. In the crystal
electric field the strongly localized $S=2$ high spin states of
 $\mathrm{Fe}^{2+}$ form two orthogonal 2D-Ising-like magnetic sublattices along O-Fe-O bonds. 
From the observed magnetic order we conclude AFM interaction via oxygen $J_\mathrm{nnn1}$ and FM interaction via
selenium $J_\mathrm{nnn2}$. \OX is the second compound along with \SrFFeS
were this particular order is observed.

\section{Acknowledgments}
This work has been financially supported by the Deutsche Forschungsgemeinschaft (DFG) 
through the priority programm SPP 1458 (projects KL~1086/10-1 and BU~887/15-1) 
and the Research Training Group GRK 1621. 
R. Sarkar is thankful to DFG for the financial support with grant no. DFG SA~2426/1-1.
This work was supported by Korea NRF Grants (No. 2012-046138 and No. 2012M7A1A2055645).
Part of this work was performed at the Swiss Muon Source (Villigen, Switzerland).

\bibliography{LaFeSeO-Draft}

\begin{thebibliography}{26}
\expandafter\ifx\csname natexlab\endcsname\relax\def\natexlab#1{#1}\fi
\expandafter\ifx\csname bibnamefont\endcsname\relax
  \def\bibnamefont#1{#1}\fi
\expandafter\ifx\csname bibfnamefont\endcsname\relax
  \def\bibfnamefont#1{#1}\fi
\expandafter\ifx\csname citenamefont\endcsname\relax
  \def\citenamefont#1{#1}\fi
\expandafter\ifx\csname url\endcsname\relax
  \def\url#1{\texttt{#1}}\fi
\expandafter\ifx\csname urlprefix\endcsname\relax\def\urlprefix{URL }\fi
\providecommand{\bibinfo}[2]{#2}
\providecommand{\eprint}[2][]{\url{#2}}

\bibitem[{\citenamefont{de~la Cruz et~al.}(2008)\citenamefont{de~la Cruz,
  Huang, Lynn, Li, II, Zarestky, Mook, Chen, Luo, Wang
  et~al.}}]{deLaCruz-Nature-2008}
\bibinfo{author}{\bibfnamefont{C.}~\bibnamefont{de~la Cruz}},
  \bibinfo{author}{\bibfnamefont{Q.}~\bibnamefont{Huang}},
  \bibinfo{author}{\bibfnamefont{J.~W.} \bibnamefont{Lynn}},
  \bibinfo{author}{\bibfnamefont{J.}~\bibnamefont{Li}},
  \bibinfo{author}{\bibfnamefont{W.~R.} \bibnamefont{II}},
  \bibinfo{author}{\bibfnamefont{J.~L.} \bibnamefont{Zarestky}},
  \bibinfo{author}{\bibfnamefont{H.~A.} \bibnamefont{Mook}},
  \bibinfo{author}{\bibfnamefont{G.~F.} \bibnamefont{Chen}},
  \bibinfo{author}{\bibfnamefont{J.~L.} \bibnamefont{Luo}},
  \bibinfo{author}{\bibfnamefont{N.~L.} \bibnamefont{Wang}},
  \bibnamefont{et~al.}, \bibinfo{journal}{Nature (London)}
  \textbf{\bibinfo{volume}{453}}, \bibinfo{pages}{899} (\bibinfo{year}{2008}),
  \urlprefix\url{http://dx.doi.org/10.1038/nature07057}.

\bibitem[{\citenamefont{Klauss et~al.}(2008)\citenamefont{Klauss, Luetkens,
  Klingeler, Hess, Litterst, Kraken, Korshunov, Eremin, Drechsler, Khasanov
  et~al.}}]{Klauss-PRL-2008}
\bibinfo{author}{\bibfnamefont{H.-H.} \bibnamefont{Klauss}},
  \bibinfo{author}{\bibfnamefont{H.}~\bibnamefont{Luetkens}},
  \bibinfo{author}{\bibfnamefont{R.}~\bibnamefont{Klingeler}},
  \bibinfo{author}{\bibfnamefont{C.}~\bibnamefont{Hess}},
  \bibinfo{author}{\bibfnamefont{F.~J.} \bibnamefont{Litterst}},
  \bibinfo{author}{\bibfnamefont{M.}~\bibnamefont{Kraken}},
  \bibinfo{author}{\bibfnamefont{M.~M.} \bibnamefont{Korshunov}},
  \bibinfo{author}{\bibfnamefont{I.}~\bibnamefont{Eremin}},
  \bibinfo{author}{\bibfnamefont{S.-L.} \bibnamefont{Drechsler}},
  \bibinfo{author}{\bibfnamefont{R.}~\bibnamefont{Khasanov}},
  \bibnamefont{et~al.}, \bibinfo{journal}{Phys. Rev. Lett.}
  \textbf{\bibinfo{volume}{101}}, \bibinfo{pages}{077005}
  (\bibinfo{year}{2008}),
  \urlprefix\url{http://link.aps.org/doi/10.1103/PhysRevLett.101.077005}.

\bibitem[{\citenamefont{Rotter et~al.}(2008)\citenamefont{Rotter, Tegel,
  Johrendt, Schellenberg, Hermes, and P\"ottgen}}]{Rotter-PRB-2008}
\bibinfo{author}{\bibfnamefont{M.}~\bibnamefont{Rotter}},
  \bibinfo{author}{\bibfnamefont{M.}~\bibnamefont{Tegel}},
  \bibinfo{author}{\bibfnamefont{D.}~\bibnamefont{Johrendt}},
  \bibinfo{author}{\bibfnamefont{I.}~\bibnamefont{Schellenberg}},
  \bibinfo{author}{\bibfnamefont{W.}~\bibnamefont{Hermes}}, \bibnamefont{and}
  \bibinfo{author}{\bibfnamefont{R.}~\bibnamefont{P\"ottgen}},
  \bibinfo{journal}{Phys. Rev. B} \textbf{\bibinfo{volume}{78}},
  \bibinfo{pages}{020503} (\bibinfo{year}{2008}),
  \urlprefix\url{http://link.aps.org/doi/10.1103/PhysRevB.78.020503}.

\bibitem[{\citenamefont{Mazin}(2010)}]{Mazin-Nature-2010}
\bibinfo{author}{\bibfnamefont{I.~I.} \bibnamefont{Mazin}},
  \bibinfo{journal}{Nature (London)} \textbf{\bibinfo{volume}{464}},
  \bibinfo{pages}{183} (\bibinfo{year}{2010}),
  \urlprefix\url{http://dx.doi.org/10.1038/nature08914}.

\bibitem[{\citenamefont{Si}(2009)}]{Si-NatPhys-2009}
\bibinfo{author}{\bibfnamefont{Q.}~\bibnamefont{Si}}, \bibinfo{journal}{Nat.
  Phys.} \textbf{\bibinfo{volume}{5}}, \bibinfo{pages}{629}
  (\bibinfo{year}{2009}), \urlprefix\url{http://dx.doi.org/10.1038/nphys1394}.

\bibitem[{\citenamefont{Fernandes et~al.}(2012)\citenamefont{Fernandes,
  Chubukov, Knolle, Eremin, and Schmalian}}]{Fernandes-PRB-2012}
\bibinfo{author}{\bibfnamefont{R.~M.} \bibnamefont{Fernandes}},
  \bibinfo{author}{\bibfnamefont{A.~V.} \bibnamefont{Chubukov}},
  \bibinfo{author}{\bibfnamefont{J.}~\bibnamefont{Knolle}},
  \bibinfo{author}{\bibfnamefont{I.}~\bibnamefont{Eremin}}, \bibnamefont{and}
  \bibinfo{author}{\bibfnamefont{J.}~\bibnamefont{Schmalian}},
  \bibinfo{journal}{Phys. Rev. B} \textbf{\bibinfo{volume}{85}},
  \bibinfo{pages}{024534} (\bibinfo{year}{2012}),
  \urlprefix\url{http://link.aps.org/doi/10.1103/PhysRevB.85.024534}.

\bibitem[{\citenamefont{Ni et~al.}(2010)\citenamefont{Ni, Climent-Pascual, Jia,
  Huang, and Cava}}]{Ni-2010-PRB}
\bibinfo{author}{\bibfnamefont{N.}~\bibnamefont{Ni}},
  \bibinfo{author}{\bibfnamefont{E.}~\bibnamefont{Climent-Pascual}},
  \bibinfo{author}{\bibfnamefont{S.}~\bibnamefont{Jia}},
  \bibinfo{author}{\bibfnamefont{Q.}~\bibnamefont{Huang}}, \bibnamefont{and}
  \bibinfo{author}{\bibfnamefont{R.~J.} \bibnamefont{Cava}},
  \bibinfo{journal}{Phys. Rev. B} \textbf{\bibinfo{volume}{82}},
  \bibinfo{pages}{214419} (\bibinfo{year}{2010}),
  \urlprefix\url{http://link.aps.org/doi/10.1103/PhysRevB.82.214419}.

\bibitem[{\citenamefont{Zhao et~al.}(2013)\citenamefont{Zhao, Wu, Wang, Hodges,
  Broholm, and Morosan}}]{Zhao-PRB-2013}
\bibinfo{author}{\bibfnamefont{L.~L.} \bibnamefont{Zhao}},
  \bibinfo{author}{\bibfnamefont{S.}~\bibnamefont{Wu}},
  \bibinfo{author}{\bibfnamefont{J.~K.} \bibnamefont{Wang}},
  \bibinfo{author}{\bibfnamefont{J.}~\bibnamefont{Hodges}},
  \bibinfo{author}{\bibfnamefont{C.}~\bibnamefont{Broholm}}, \bibnamefont{and}
  \bibinfo{author}{\bibfnamefont{E.}~\bibnamefont{Morosan}},
  \bibinfo{journal}{Phys. Rev. B} \textbf{\bibinfo{volume}{87}},
  \bibinfo{pages}{020406(R)} (\bibinfo{year}{2013}),
  \urlprefix\url{http://link.aps.org/doi/10.1103/PhysRevB.87.020406}.

\bibitem[{\citenamefont{Fuwa et~al.}(2010{\natexlab{a}})\citenamefont{Fuwa,
  Endo, Wakeshima, Hinatsu, and Ohoyama}}]{Fuwa-JACS-2010}
\bibinfo{author}{\bibfnamefont{Y.}~\bibnamefont{Fuwa}},
  \bibinfo{author}{\bibfnamefont{T.}~\bibnamefont{Endo}},
  \bibinfo{author}{\bibfnamefont{M.}~\bibnamefont{Wakeshima}},
  \bibinfo{author}{\bibfnamefont{Y.}~\bibnamefont{Hinatsu}}, \bibnamefont{and}
  \bibinfo{author}{\bibfnamefont{K.}~\bibnamefont{Ohoyama}},
  \bibinfo{journal}{J. Am. Chem. Soc.} \textbf{\bibinfo{volume}{132}},
  \bibinfo{pages}{18020} (\bibinfo{year}{2010}{\natexlab{a}}),
  \urlprefix\url{http://pubs.acs.org/doi/abs/10.1021/ja109007g}.

\bibitem[{\citenamefont{Mayer et~al.}(1992)\citenamefont{Mayer, Schneemeyer,
  Siegrist, Waszczak, and Van~Dover}}]{Mayer1992}
\bibinfo{author}{\bibfnamefont{J.~M.} \bibnamefont{Mayer}},
  \bibinfo{author}{\bibfnamefont{L.~F.} \bibnamefont{Schneemeyer}},
  \bibinfo{author}{\bibfnamefont{T.}~\bibnamefont{Siegrist}},
  \bibinfo{author}{\bibfnamefont{J.~V.} \bibnamefont{Waszczak}},
  \bibnamefont{and}
  \bibinfo{author}{\bibfnamefont{B.}~\bibnamefont{Van~Dover}},
  \bibinfo{journal}{Angew. Chem., Int. Ed. Engl.}
  \textbf{\bibinfo{volume}{31}}, \bibinfo{pages}{1645} (\bibinfo{year}{1992}),
  \urlprefix\url{http://dx.doi.org/10.1002/anie.199216451}.

\bibitem[{\citenamefont{Kabbour et~al.}(2008)\citenamefont{Kabbour, Janod,
  Corraze, Danot, Lee, Whangbo, and Cario}}]{Kabbour-JACS-2008}
\bibinfo{author}{\bibfnamefont{H.}~\bibnamefont{Kabbour}},
  \bibinfo{author}{\bibfnamefont{E.}~\bibnamefont{Janod}},
  \bibinfo{author}{\bibfnamefont{B.}~\bibnamefont{Corraze}},
  \bibinfo{author}{\bibfnamefont{M.}~\bibnamefont{Danot}},
  \bibinfo{author}{\bibfnamefont{C.}~\bibnamefont{Lee}},
  \bibinfo{author}{\bibfnamefont{M.-H.} \bibnamefont{Whangbo}},
  \bibnamefont{and} \bibinfo{author}{\bibfnamefont{L.}~\bibnamefont{Cario}},
  \bibinfo{journal}{J. Am. Chem. Soc.} \textbf{\bibinfo{volume}{130}},
  \bibinfo{pages}{8261} (\bibinfo{year}{2008}),
  \urlprefix\url{http://dx.doi.org/10.1021/ja711139g}.

\bibitem[{\citenamefont{Fuwa et~al.}(2010{\natexlab{b}})\citenamefont{Fuwa,
  Wakeshima, and Hinatsu}}]{Fuwa-JPCM-2010}
\bibinfo{author}{\bibfnamefont{Y.}~\bibnamefont{Fuwa}},
  \bibinfo{author}{\bibfnamefont{M.}~\bibnamefont{Wakeshima}},
  \bibnamefont{and} \bibinfo{author}{\bibfnamefont{Y.}~\bibnamefont{Hinatsu}},
  \bibinfo{journal}{J. Phys.: Condens. Matter} \textbf{\bibinfo{volume}{22}},
  \bibinfo{pages}{346003} (\bibinfo{year}{2010}{\natexlab{b}}),
  \urlprefix\url{http://stacks.iop.org/0953-8984/22/i=34/a=346003}.

\bibitem[{\citenamefont{Free et~al.}(2011)\citenamefont{Free, Withers, Hickey,
  and Evans}}]{Free2011}
\bibinfo{author}{\bibfnamefont{D.~G.} \bibnamefont{Free}},
  \bibinfo{author}{\bibfnamefont{N.~D.} \bibnamefont{Withers}},
  \bibinfo{author}{\bibfnamefont{P.~J.} \bibnamefont{Hickey}},
  \bibnamefont{and} \bibinfo{author}{\bibfnamefont{J.~S.~O.}
  \bibnamefont{Evans}}, \bibinfo{journal}{Chem. Mater.}
  \textbf{\bibinfo{volume}{23}}, \bibinfo{pages}{1625} (\bibinfo{year}{2011}),
  \urlprefix\url{http://pubs.acs.org/doi/abs/10.1021/cm1035453}.

\bibitem[{\citenamefont{Landsgesell et~al.}(2013)\citenamefont{Landsgesell,
  Blumenr\"other, and Proke\v{s}}}]{Landsgesell2013}
\bibinfo{author}{\bibfnamefont{S.}~\bibnamefont{Landsgesell}},
  \bibinfo{author}{\bibfnamefont{E.}~\bibnamefont{Blumenr\"other}},
  \bibnamefont{and}
  \bibinfo{author}{\bibfnamefont{K.}~\bibnamefont{Proke\v{s}}},
  \bibinfo{journal}{J. Phys.: Condens. Matter} \textbf{\bibinfo{volume}{25}},
  \bibinfo{pages}{086004} (\bibinfo{year}{2013}),
  \urlprefix\url{http://stacks.iop.org/0953-8984/25/i=8/a=086004}.

\bibitem[{\citenamefont{Lei et~al.}(2012)\citenamefont{Lei, Bozin, Llobet,
  Ivanovski, Koteski, Belosevic-Cavor, Cekic, and Petrovic}}]{Lei-PRB-2012}
\bibinfo{author}{\bibfnamefont{H.}~\bibnamefont{Lei}},
  \bibinfo{author}{\bibfnamefont{E.~S.} \bibnamefont{Bozin}},
  \bibinfo{author}{\bibfnamefont{A.}~\bibnamefont{Llobet}},
  \bibinfo{author}{\bibfnamefont{V.}~\bibnamefont{Ivanovski}},
  \bibinfo{author}{\bibfnamefont{V.}~\bibnamefont{Koteski}},
  \bibinfo{author}{\bibfnamefont{J.}~\bibnamefont{Belosevic-Cavor}},
  \bibinfo{author}{\bibfnamefont{B.}~\bibnamefont{Cekic}}, \bibnamefont{and}
  \bibinfo{author}{\bibfnamefont{C.}~\bibnamefont{Petrovic}},
  \bibinfo{journal}{Phys. Rev. B} \textbf{\bibinfo{volume}{86}},
  \bibinfo{pages}{125122} (\bibinfo{year}{2012}),
  \urlprefix\url{http://link.aps.org/doi/10.1103/PhysRevB.86.125122}.

\bibitem[{\citenamefont{Zhu et~al.}(2010)\citenamefont{Zhu, Yu, Wang, Zhao,
  Jones, Dai, Abrahams, Morosan, Fang, and Si}}]{Zhu-2010}
\bibinfo{author}{\bibfnamefont{J.-X.} \bibnamefont{Zhu}},
  \bibinfo{author}{\bibfnamefont{R.}~\bibnamefont{Yu}},
  \bibinfo{author}{\bibfnamefont{H.}~\bibnamefont{Wang}},
  \bibinfo{author}{\bibfnamefont{L.~L.} \bibnamefont{Zhao}},
  \bibinfo{author}{\bibfnamefont{M.~D.} \bibnamefont{Jones}},
  \bibinfo{author}{\bibfnamefont{J.}~\bibnamefont{Dai}},
  \bibinfo{author}{\bibfnamefont{E.}~\bibnamefont{Abrahams}},
  \bibinfo{author}{\bibfnamefont{E.}~\bibnamefont{Morosan}},
  \bibinfo{author}{\bibfnamefont{M.}~\bibnamefont{Fang}}, \bibnamefont{and}
  \bibinfo{author}{\bibfnamefont{Q.}~\bibnamefont{Si}}, \bibinfo{journal}{Phys.
  Rev. Lett.} \textbf{\bibinfo{volume}{104}}, \bibinfo{pages}{216405}
  (\bibinfo{year}{2010}),
  \urlprefix\url{http://link.aps.org/doi/10.1103/PhysRevLett.104.216405}.

\bibitem[{\citenamefont{Wang et~al.}(2011)\citenamefont{Wang, Zhang, Zheng, and
  Yang}}]{Wang-2011-SSC}
\bibinfo{author}{\bibfnamefont{G.}~\bibnamefont{Wang}},
  \bibinfo{author}{\bibfnamefont{M.}~\bibnamefont{Zhang}},
  \bibinfo{author}{\bibfnamefont{L.}~\bibnamefont{Zheng}}, \bibnamefont{and}
  \bibinfo{author}{\bibfnamefont{Z.}~\bibnamefont{Yang}},
  \bibinfo{journal}{Solid State Communications} \textbf{\bibinfo{volume}{151}},
  \bibinfo{pages}{1231 } (\bibinfo{year}{2011}),
  \urlprefix\url{http://www.sciencedirect.com/science/article/pii/S0038109811002924}.

\bibitem[{\citenamefont{Free and Evans}(2010)}]{Free-2010-PRB}
\bibinfo{author}{\bibfnamefont{D.~G.} \bibnamefont{Free}} \bibnamefont{and}
  \bibinfo{author}{\bibfnamefont{J.~S.~O.} \bibnamefont{Evans}},
  \bibinfo{journal}{Phys. Rev. B} \textbf{\bibinfo{volume}{81}},
  \bibinfo{pages}{214433} (\bibinfo{year}{2010}),
  \urlprefix\url{http://link.aps.org/doi/10.1103/PhysRevB.81.214433}.

\bibitem[{\citenamefont{McCabe et~al.}(2014)\citenamefont{McCabe, Stock,
  Rodriguez, Wills, Taylor, and Evans}}]{McCabe-PRBR-2014}
\bibinfo{author}{\bibfnamefont{E.~E.} \bibnamefont{McCabe}},
  \bibinfo{author}{\bibfnamefont{C.}~\bibnamefont{Stock}},
  \bibinfo{author}{\bibfnamefont{E.~E.} \bibnamefont{Rodriguez}},
  \bibinfo{author}{\bibfnamefont{A.~S.} \bibnamefont{Wills}},
  \bibinfo{author}{\bibfnamefont{J.~W.} \bibnamefont{Taylor}},
  \bibnamefont{and} \bibinfo{author}{\bibfnamefont{J.~S.~O.}
  \bibnamefont{Evans}}, \bibinfo{journal}{Phys. Rev. B}
  \textbf{\bibinfo{volume}{89}}, \bibinfo{pages}{100402}
  (\bibinfo{year}{2014}),
  \urlprefix\url{http://link.aps.org/doi/10.1103/PhysRevB.89.100402}.

\bibitem[{\citenamefont{Rietveld}(1969)}]{Rietveld-JApplCryst-1969}
\bibinfo{author}{\bibfnamefont{H.~M.} \bibnamefont{Rietveld}},
  \bibinfo{journal}{J. Appl. Crystallogr.} \textbf{\bibinfo{volume}{2}},
  \bibinfo{pages}{65} (\bibinfo{year}{1969}),
  \urlprefix\url{http://dx.doi.org/10.1107/S0021889869006558}.

\bibitem[{\citenamefont{Roisnel and
  Rodriguez-Carvajal}(2001)}]{Roisnel-MatrSciForum-2001}
\bibinfo{author}{\bibfnamefont{T.}~\bibnamefont{Roisnel}} \bibnamefont{and}
  \bibinfo{author}{\bibfnamefont{J.}~\bibnamefont{Rodriguez-Carvajal}},
  \bibinfo{journal}{Mater. Sci. Forum} \textbf{\bibinfo{volume}{378-381}},
  \bibinfo{pages}{118} (\bibinfo{year}{2001}),
  \urlprefix\url{http://www.scientific.net/MSF.378-381.118}.

\bibitem[{\citenamefont{Suter and Wojek}(2012)}]{Suter-PhysProcedia-2012}
\bibinfo{author}{\bibfnamefont{A.}~\bibnamefont{Suter}} \bibnamefont{and}
  \bibinfo{author}{\bibfnamefont{B.~M.} \bibnamefont{Wojek}},
  \bibinfo{journal}{Phys. Procedia} \textbf{\bibinfo{volume}{30}},
  \bibinfo{pages}{69} (\bibinfo{year}{2012}),
  \urlprefix\url{http://dx.doi.org/10.1016/j.phpro.2012.04.042}.

\bibitem[{\citenamefont{Shirane et~al.}(1962)\citenamefont{Shirane, Cox, and
  Ruby}}]{Shirane-PR-1962}
\bibinfo{author}{\bibfnamefont{G.}~\bibnamefont{Shirane}},
  \bibinfo{author}{\bibfnamefont{D.~E.} \bibnamefont{Cox}}, \bibnamefont{and}
  \bibinfo{author}{\bibfnamefont{S.~L.} \bibnamefont{Ruby}},
  \bibinfo{journal}{Phys. Rev.} \textbf{\bibinfo{volume}{125}},
  \bibinfo{pages}{1158} (\bibinfo{year}{1962}),
  \urlprefix\url{http://link.aps.org/doi/10.1103/PhysRev.125.1158}.

\bibitem[{\citenamefont{Moriya}(1956)}]{Moriya-PTP-1956}
\bibinfo{author}{\bibfnamefont{T.}~\bibnamefont{Moriya}},
  \bibinfo{journal}{Prog. Theor. Phys.} \textbf{\bibinfo{volume}{16}},
  \bibinfo{pages}{23} (\bibinfo{year}{1956}),
  \urlprefix\url{http://dx.doi.org/10.1143/PTP.16.23}.

\bibitem[{\citenamefont{Beeman and Pincus}(1968)}]{Beeman-PhysRev-1968}
\bibinfo{author}{\bibfnamefont{D.}~\bibnamefont{Beeman}} \bibnamefont{and}
  \bibinfo{author}{\bibfnamefont{P.}~\bibnamefont{Pincus}},
  \bibinfo{journal}{Phys. Rev.} \textbf{\bibinfo{volume}{166}},
  \bibinfo{pages}{359} (\bibinfo{year}{1968}),
  \urlprefix\url{http://journals.aps.org/pr/pdf/10.1103/PhysRev.166.359}.

\bibitem[{\citenamefont{de~R\'eotier and Yaouanc}(1997)}]{muSR}
\bibinfo{author}{\bibfnamefont{P.~D.} \bibnamefont{de~R\'eotier}}
  \bibnamefont{and} \bibinfo{author}{\bibfnamefont{A.}~\bibnamefont{Yaouanc}},
  \bibinfo{journal}{J. Phys.: Condens. Matter} \textbf{\bibinfo{volume}{9}},
  \bibinfo{pages}{9113} (\bibinfo{year}{1997}),
  \urlprefix\url{http://stacks.iop.org/0953-8984/9/i=43/a=002}.

\end{thebibliography}

\end{document}